\documentclass[notitlepage, article,amsmath,amssymb,aps, onecolumn]{revtex4-1}
\DeclareMathOperator{\arccosh}{arccosh}
\DeclareMathOperator{\arcsinh}{arcsinh}
\DeclareMathOperator{\arctanh}{arctanh}
\usepackage{graphicx}
\usepackage{dcolumn}
\usepackage{bm}

\begin{document}

\title{The electromagnetic self-force of a Lorentz-contractible spherical shell of radius $R$ in rectilinear arbitrary motion: the terms of order $1/R$ and $R^0$}

\author{G. Vaman\\
 Institute of Atomic Physics, P. O. Box MG-6, Bucharest, Romania}
 \email{ getavaman@gmail.com}

\date{\today}

\begin{abstract}
We write the electromagnetic self-force of a Lorentz-contractible spherical shell of radius $R$ in arbitrary rectilinear motion as a series expansion in powers of $R$, and calculate the first two terms of this series. The method we use, which is based on the description of the particle in terms of a velocity-dependent volume charge density, avoids some important difficulties of the previous approaches from the literature. We compare our results with the results obtained by other authors. The whole calculation is done in the laboratory frame of reference.
\end{abstract}

\pacs{Valid PACS appear here}
\maketitle


\section{Introduction}
The recent development of high-power lasers made possible the experimental study of the interaction of the relativistic charged particles with external electromagnetic fields \cite{col} - \cite{had}. Comparing the experimental data with the results given by various theoretical models makes it possible to clearly establish the  applicability limits of the later.

The theoretical study of the electromagnetic self-force acting on a moving charged object has a long history \cite{mcd}. A charged object moving under the action of an external force radiates, so it loses energy. Equivalently, one can says that a braking force acts on it, due to the action of electromagnetic field radiated by the body on itself. The calculation of this self-force can be approached either by invoking the conservation of  momentum for the physical system consisting of the charged body and the electromagnetic field that surrounds it, or by integrating the Lorentz force density of the self-field on the considered object \cite{abr} - \cite{vam}.

When it comes to the evaluation of the electromagnetic self-force for charged extended bodies, leaving aside the problems connected with the stability of electric charges  and with the propagation of the forces inside the moving charged particles, it is still not easy to do the calculations so that the principles of the theory of relativity are satisfied. Usually, the computation is done by considering the extendend body as being made up of infinitesimal charges that act on each other and adding up (integrating) all these forces. Considering that Born rigidity condition is satisfied, which means that the charged body has a constant shape and charge density in a momentarily co-moving inertial frame (proper instantaneous frame), as noted in \cite{yag}, we have to take into account the fact that, in the laboratory frame of reference,  the speed of an infinitesimal charge inside the particle is generally not equal to the speed of the particle's center. Moreover, even in the instantaneous proper frame, the acceleration of an infinitesimal charge and all the higher derivatives of its velocity  depend on the position of the infinitesimal charge inside the extended body \cite{yag}, \cite{lyl}. Evidently, this dependence can be taken into account only approximately.

 We present in this paper a new method of calculating the electromagnetic self-force of a Lorentz-contractible charged spherical shell. This method avoids some important difficulties that arise when we consider separately the contribution of each elementary charge to the total self-force, as we show in Sec. VI. We consider the shell as being described by a velocity-dependent volume charge density so that a shell that has a spherical shape in its instantaneous proper frame will have the shape of an oblate spheroid, whose dimensions depend on velocity, in the laboratory frame. Also, while the charge density of the shell is constant in its instantaneous proper frame, it becomes dependent on the position in the laboratory frame \cite{tor}. Although we cannot yet calculate the exact expression of the self-force of a Lorentz-contractible shell, we are able to write it as a series expansion in powers of its characteristic length and calculate the first terms of this expansion. 

In the second section of this paper we present our method of calculation, in Secs. III - V we calculate the first two non-zero terms in the expansion of the electromagnetic self-force and in the end section we discuss our results and compare them with other results from the literature. Some longer mathematical justifications of the formulas appearing in this paper are presented in detail in the Appendices.

\section{The method of calculation}
We consider a shell that, in its instantaneously proper frame of reference, has a spherical shape of radius $R$ and it is uniformly charged with a charge density $Q/(4\pi R^2)$. In the laboratory frame, this shell is moving along the $x$-axis, and we denote by $w(t) {\bf i}$ the trajectory of its center, where {\bf i} is the unit vector along the $x$-axis. For simplicity, we consider that, at the current time $t$, we have $w(t)=0$.  In the laboratory frame of reference, the shell is Lorentz-contracted in the direction of the $x$-axis, having the shape of an oblate spheroid of semi-axes $\left( \frac{R}{\gamma},R,R \right)$, where $ \gamma=\frac{1}{\sqrt{1-\beta^2}}$, $ \beta =\dot{w}(t)/c$, and has a new charge density which ensures the equilibrium of the electrical charges on its surface \cite{tor}. 
As our calculation is performed in the laboratory frame of reference, it is convenient to use the oblate spheroidal coordinates $\eta, \theta, \phi$ \cite{mor}
\begin{align}
&x-w(t)= - P \sinh  \eta \cos \theta\nonumber \\
&y= P \cosh \eta \sin \theta \cos \phi\\
&z= P \cosh \eta \sin \theta \sin \phi, \nonumber
\end{align}
where $\eta \in [0, \infty)$, $\theta \in [0, \pi)$, $\phi \in [0, 2\pi)$ and $P$ is an arbitrary parameter,
and to choose the parameter $P$ so that one of the  coordinate surfaces $\eta=\mbox{constant}$ coincides with the moving charged shell. Choosing $P = R \beta$, the surface $\eta=\eta_0=\arcsinh \frac{1}{\beta \gamma}= \arccosh \frac{1}{\beta}$ is an oblate spheroid of semi-axes $\left(\frac{R}{\gamma},R,R\right)$  centered at the point $(w(t),0,0)$, so it coincides with the surface of the Lorentz-contracted  shell. Thus, the points on the moving shell are given by
\begin{align}
&x= w(t)- \frac{R}{\gamma} \cos \theta\nonumber \\
&y=R  \sin \theta \cos \phi\\
&z=R  \sin \theta \sin \phi. \nonumber
\end{align}
Although, for shortness, we sometimes drop the argument $t$ of $\gamma(t)$, we have to keep in mind that everywhere in this paper $\gamma$ is a time-dependent quantity.
To perform the integrals, we need the volume element in oblate spheroidal coordinates that can be obtained from the Lam\'{e} coefficients
\begin{align}
&h_{\eta}= h_{\theta}= R \beta \sqrt{\sinh^2\eta+\cos^2 \theta}\nonumber \\
&h_{\phi}= R \beta \cosh \eta \sin \theta\nonumber
\end{align}
as
\begin{align}
d {\bf r}=h_\eta h_\theta h_\phi = &R^3 \beta^3 \cosh \eta \sin \theta 
 (\sinh^2 \eta + \cos^2 \theta) d \eta d \theta d \phi.
\end{align}
It has been proved in \cite{cur} that the volume charge density of an ellipsoid of semi-axes $(a,b,c)$ can be written as 
\begin{align}
\rho({\bf r})= \frac{Q}{4\pi abc}\delta \left( \sqrt{\frac{x^2}{a^2}+ \frac{y^2}{b^2} +\frac{z^2}{c^2}}-1   \right),
\end{align}
and that from this charge density one obtains the well-known surface charge density of an ellipsoid \cite{tor}
\begin{align}
\sigma({\bf r})= \frac{Q}{4\pi abc} \left( \frac{x^2}{a^4} +\frac{y^2}{b^4} +\frac{z^2}{c^4} \right)^{-1/2}_{|\mbox{x,y,z on the ellipsoid}}.
\end{align}
Using the result from \cite{cur}, we write the volume charge density of our charged shell whose center is at the point $(w(t),0,0)$ as
\begin{align}
\rho({\bf r},t)= \frac{Q \gamma}{4\pi R^3} \delta \left( \sqrt{ \frac{(x-w(t))^2}{a^2}+\frac{y^2}{b^2} + \frac{z^2}{c^2}}-1 \right)=\frac{Q \gamma}{4\pi R^3} \sum_{p=0}^{\infty} \frac{(-1)^pw^p(t)}{p!} \partial_x^p \delta \left( \sqrt{ \frac{x^2}{a^2}+\frac{y^2}{b^2} + \frac{z^2}{c^2}}-1 \right),
\end{align}
where, in our case, $a=R/\gamma$, $b=c=R$, and the series expansion in powers of $w(t)$ is convenient for performing the necessary integrals. After we 
 perform all the integrals, we will also perform the summation over $p$,  so our result will be valid for an arbitrary motion along the $x$-axis, not only for small oscillations $w(t)$ around the origin.

Writing $x,y,z$ in terms of oblate spheroidal coordinates as in Eq.(1) and using the property of the Dirac delta function
\begin{align}
\delta(f(\eta))=\frac{\delta(\eta-\eta_0)}{|f'(\eta_0)|},
\end{align}
where $f(\eta_0)=0$, we obtain
\begin{align}
\delta \left( \sqrt{ \frac{x^2}{a^2}+\frac{y^2}{b^2} + \frac{z^2}{c^2}}-1 \right)= \frac{\gamma \delta(\eta-\eta_0)}{\gamma^2\cos^2\theta +\sin^2 \theta}.
\end{align}

The electromagnetic self-force acting on the charged ellipsoid can be written in terms of the Lorentz force density
\begin{align}\label{1}
{\bf F}=\int d {\bf r} \left( \rho( {\bf r},t) {\bf E}_s   ( {\bf r},t) +\frac{1}{c} {\bf J} ( {\bf r},t) \times  {\bf B}_s   ( {\bf r},t)\right),
\end{align}
where the self-fields are given in terms of the electromagnetic potentials as ${\bf E}_s   ( {\bf r},t) =- \nabla \phi( {\bf r},t)-\frac{1}{c} \frac{\partial {\bf A}( {\bf r},t)}{\partial t}$, ${\bf B}_s   ( {\bf r},t)= \nabla \times {\bf A}  ( {\bf r},t)$.
Next, following the method from \cite{jac}, we write the electromagnetic potentials in     terms of retarded charge and current densities $\phi ( {\bf r},t) =\int d {\bf r'} \frac{\rho({\bf r'},t')}{R}$, ${\bf A} ( {\bf r},t) =\frac{1}{c}\int d {\bf r'} \frac{{\bf J}({\bf r'},t')}{R} $, $R=|{\bf r}-{\bf r'}|$, $t'=t-R/c$, and expand the retarded charge and current densities in Taylor series around $t'=t$. After noting that, in the case of rectilinear motion, the second term in the r.h.s. of Eq. (9) gives no contribution because the vectors ${\bf J}$ and ${\bf B_s}$ are orthogonal, we obtain
\begin{align}
{\bf F}(t)={\bf F^{(1)}}(t)+{\bf F^{(2)}}(t), \label{2}
\end{align}
where
\begin{align}\label{3}
{\bf F^{(1)}}(t)=\sum_{n=0}^{\infty} \frac{(-1)^{n+1}}{n!\,c^{n+2}} \frac{n+1}{n+2} \int d {\bf r} \int d {\bf r'} \rho({\bf r'},t) | {\bf r}- {\bf r'}|^{n-1} 
  \frac{\partial^{n+1}}{\partial t^{n+1}} \left( \rho({\bf r},t) {\bf v}(t) \right),
\end{align}
\begin{align}\label{4}
{\bf F^{(2)}}(t) = \sum_{n=0}^{\infty} \frac{(-1)^{n}}{n!\,c^{n+2}} \frac{n-1}{n+2} \int d {\bf r} \int d {\bf r'}  \rho({\bf r'},t) | {\bf r}- {\bf r'}|^{n-3}  \frac{\partial^{n+1}}{\partial t^{n+1}} \left( \rho({\bf r},t) {\bf v}(t) \cdot ({\bf r}- {\bf r'}) ({\bf r}- {\bf r'})\right).
\end{align}
The quantities ${\bf F^{(1)}}(t)$ and ${\bf F^{(2)}}(t)$ from Eqs. (11), (12) are the first and the second term of Eqs. (16.25) + (16.26) from \cite{jac} with changed sign, respectively.\\

\noindent
Using our Eqs. (3), (8), all the spatial integrals can be written as follows
\begin{align}
&\int d{\bf r} \;\delta \left( \sqrt{ \frac{x^2}{a^2}+\frac{y^2}{b^2} + \frac{z^2}{c^2}}-1 \right) f({\bf r})=\frac{R^3}{\gamma}\int_0^{\pi}d \theta \int_0^{2\pi} d \phi \;\sin \theta \;f(\eta_0,\theta,\phi),
\end{align}
where $f$ is an arbitrary function.

For the inverse distance between two points on the spheroid raised to an arbitrary power, we use (see Appendix A)
\begin{align}
&\frac{1}{|{\bf r}-{\bf r'}|_{\text{sph}}^n} = \frac{1}{R^n 2^{n/2}} \sum_{r=0}^{\infty} \sum_{k=0}^{\infty} \sum_{q=0}^{[\frac{k}{2}]} \sum_{m=-k+2q}^{k+2q} \frac{\beta^{2r} \Gamma \left( r+ \frac{n}{2} \right)}{2^r r!\Gamma \left( \frac{n}{2}\right) }  (\cos \theta - \cos \theta')^{2r} \frac{\left( k+r + \frac{n}{2}-1 \right)!}{\left( r +\frac{n}{2}-1 \right)!} \frac{(2k-4q+1)}{2^q q! (2k-2q+1)!!} \nonumber \\
&\cdot \frac{(k-2q-m)!}{(k-2q+m)!} P_{k-2q}^m (\cos \theta) P_{k-2q}^m (\cos \theta') e^{im (\phi - \phi')},
\end{align}
where the lower index "sph" means "on the oblate spheroid". We shall also need during this calculation the following formulas inferred in \cite{hni2}:\begin{align}
\frac{\partial^{n+1}}{\partial x^{n+1}} r^{i-1}&= 
\sum_{b=0}^{n/2}\frac{ (-1)^{b+1+n/2}2^{2b+1}(n+1)!} {(2b+1)! (n/2-b)!} \left(  \frac{1-i}{2}\right)_{b+1+n/2} r^{i-n-2b-3}  x^{2b+1},\: n\;{\rm even},\label{even}\\
\frac{\partial^{n+1}}{\partial x^{n+1}} r^{i-1}&= 
\sum_{b=0}^{(n+1)/2}\frac{ (-1)^{b+(n+1)/2}2^{2b}(n+1)!} {(2b)! ((n+1)/2-b)!} \left( \frac{1-i}{2}\right)_{b+(n+1)/2} r^{i-n-2b-2}  x^{2b},\: n\;{\rm odd}, \label{odd}
\end{align}
where $(\tfrac{1}{2}-\frac{i}{2})_{b+1+n/2}$,  etc.\, are the Pochhammer symbols (\cite{pru}, Appendix I).

From now on, the method of calculation follows closely the steps from \cite{vam}.
We introduce (6) in (10) - (12) we integrate by parts and perform the integrals over $\eta$ according to Eq. (13). Because of symmetry, only the $x$- component of the electromagnetic self force is different from zero. In the next three sections we shall calculate the first two terms of the expansions of $F_x^1(t)$ and $F_x^2(t)$, starting from Eqs. (10) - (12).

\section{Calculation of $F_x^{(1)}(t)$}
The $x$-component of Eq. (11) reads
\begin{align}
F_x^{(1)}(t)=\sum_{n=0}^{\infty} \frac{(-1)^{n+1}}{n!\,c^{n+2}} \frac{n+1}{n+2} 
 \int d {\bf r} \int d {\bf r'} \rho({\bf r'},t) | {\bf r}- {\bf r'}|^{n-1} 
\frac{\partial^{n+1}}{\partial t^{n+1}} \left( \rho({\bf r},t) \dot{w}(t) \right).
\end{align}
Using Eqs. (6), (13), the integral over ${\bf r}$ can be written as follows
\begin{align}
\int d {\bf r} \,| {\bf r}-{\bf r'}|^{n-1} \frac{\partial^{n+1}}{\partial t^{n+1}}(\rho({\bf r},t) \dot{w}(t))= \frac{Q}{4\pi} \sum_{p=0}^{\infty}\frac{1}{p!} \frac{\partial^{n+1}}{\partial t^{n+1}} \left[ w^p(t) \dot{w}(t) J_1(p,n; {\bf r'}) \right], 
\end{align}
where
\begin{align}
J_1(p,n; {\bf r'})= \int_0^{\pi} d \theta \int_0^{2\pi} d \phi \sin \theta \,\partial_x^p | {\bf r} - {\bf r'}|^{n-1}_{| \eta=\eta_0}.
\end{align}

It is important to note that $J_1(p,n,{\bf r'})$ depends on time, because $\eta=\eta_0$ implies that $x$ depends on time through $\gamma$, as one can see in our Eq. (2). However, for simplicity, we will omit the argument $t$ of $J$ throughout this paper.

From Eqs (17), (18) it follows
\begin{align}
&F_x^{(1)}(t)=\frac{Q}{4\pi} \sum_{p=0}^{\infty} \sum_{n=0}^{\infty} \frac{(-1)^{n+1}}{n! \,p! \, c^{n+2}} \frac{n+1}{n+2}  \int d {\bf r'} \rho( {\bf r'},t) \frac{\partial^{n+1}}{\partial t^{n+1}} \left[ w^p(t) \dot{w}(t) J_1(p,n; {\bf r'}) \right].
\end{align}
Taking into account the expression of $ \partial_x^p | {\bf r} - {\bf r'}|^{n-1}$ that follows from Eqs. (15), (16), it follows that we get non-zero contributions to $F_x^1$ only for even values of $p$ and $n \ge p-1$. Giving $n$ succesive values, we obtain the series expansion of $F_x^1(t)$ in powers of $R$
\begin{align}
F_x^{(1)}(t)= F_x^{(1)}(n=p-1)+ F_x^{(1)}(n=p) +F_x^{(1)}(n=p+1) +{\cal O}(R),
\end{align}
where $F_x^{(1)}(n=p-1)$, $F_x^{(1)}(n=p)$ and $F_x^{(1)}(n=p+1)$ are of the order $1/R^2$, $1/R$ and $R^0$ respectively. In the r.h.s. of Eq. (21), for simplicity, we omitted the argument $t$. Also, in the following, to make the formulas easier to read, we will often omit the argument $t$.

\subsection*{III.A Calculation of $F_x^{(1)}(n=p-1)$}
For $n=p-1$, we obtain from Eq. (20)
\begin{align}
&F_x^{(1)}(n=p-1)=\frac{Q}{4\pi} \sum_{p=0}^{\infty}  \frac{(-1)^{p}}{(p-1)! \,p! \, c^{p+1}} \frac{p}{p+1}  \int d {\bf r'} \rho( {\bf r'},t) \frac{\partial^{p}}{\partial t^{p}} \left[ w^p(t) \dot{w}(t) J_1(p,p-1; {\bf r'}) \right].\nonumber
\end{align}
As $\frac{d^p}{dt^p}w^k(t)=0$ for $p<k$, it follows that
\begin{align}
\frac{d^p}{dt^p} [w^p(t) \dot{w}(t) J_1(p,p-1,{\bf r'})]= p! \dot{w}^{p+1}(t) J_1(p,p-1,{\bf r'}). \nonumber
\end{align}
For $p$ odd, taking into account our Eqs. (15), (19), it follows that $\partial_x^p|{\bf r}- {\bf r'}|^{p-2}$ that appears in the expression of $ J_1(p,p-1,{\bf r'}) $ contains odd powers of $(x-x')$  that give no contribution to the double integral that appears in the expression of $F_x^{(1)}(n=p-1).$
For $p$ even,  
$\left( -\frac{p}{2}+1\right)_{b+\frac{p}{2}}=0, 
$  (because a Pochhammer symbol of type $(-n)_k$, where $n$ is a positive integer, is different from zero only for $k \le n)$, so it follows that 
$
\frac{\partial^p}{\partial x^p} |{\bf r}-{\bf r'}|^{p-2} \stackrel{\text{p even}}{=}0, 
$
so $ J_1(p, p-1, {\bf r}')=0 $. It follows that $F^1_x(n=p-1)= 0$.

\subsection*{III.B Calculation of $F_x^{(1)}(n=p)$}
After remembering that $\frac{d^p}{dt^p}w^k(t) \ne 0$  only for $p \ge k$, it follows that 
\begin{align}
\frac{\partial^{p+1}}{\partial t^{p+1}}(w^p\dot{w}J_1)=(p+1)! \dot{w}^{p+1} \frac{\partial J_1}{\partial t} +(p+1)! \left( 1+\frac{p}{2} \right) \dot{w}^p \ddot{w} J_1.
\end{align}
$J_1(p,p,r')$ is zero for $p$ odd because $\frac{\partial^p}{\partial x^p} |{\bf r}-{\bf r'}|^{p-1}_{|p odd}$ contains $\left( \frac{1-p}{2}\right)_{b+\frac{p+1}{2}}=0.$
From Eqs. (20), (22), one obtains
\begin{align}
F_x^{(1)}(n=p)= -\frac{Q\ddot{w}}{8\pi c^2} \sum_{\substack{p=0 \\ p \;\text{even}}}^{\infty} \frac{(p+1)^2}{p!} \beta^p \cdot A^{(1)}- \frac{Q \dot{w}}{4\pi c^2}                      
  \sum_{\substack{p=0 \\ p \;\text{even}}}^{\infty} \frac{(p+1)^2}{p! (p+2)} \beta^p \cdot C^{(1)},
\end{align}
where
\begin{align}
A^{(1)} = \int d {\bf r'} \rho({\bf r'},t) J_1(p,p,{\bf r'})
\end{align}
and
\begin{align}
C^{(1)} = \int d {\bf r'} \rho({\bf r'},t) \frac{\partial J_1(p,p,{\bf r'})}{\partial t}
\end{align}
To calculate the derivative of $J_1$ with respect to time that we need in Eq.(25), we use the chain rule for the derivative of a composite function
\begin{align}
\frac{\partial J_1}{\partial t} = \frac{\partial J_1}{\partial x} \cdot \frac{\partial x}{\partial t}.
\end{align}
Using Eqs. (15), (16), $A^{(1)}$ and $C^{(1)}$ can be written in terms of the quantities $J(b)$ and $I(b)$ from Appendix B as follows
\begin{align}
A^{(1)}=\frac{Q}{4\pi} \sum_{b=0}^{\frac{p}{2}} \frac{(-1)^{b+\frac{p}{2} }2^{2b} p!}{(2b)! \left( \frac{p}{2}-b \right)!} \left( \frac{1-p}{2} \right)_{b+\frac{p}{2}} J(b),
\end{align}
\begin{align}
C^{(1)}=\frac{Q}{4\pi} R\gamma \beta \dot{\beta} \left[ \sum_{b=0}^{\frac{p}{2}} \frac{(-1)^{b+\frac{p}{2}}2^{2b}p!}{(2b)! \left( \frac{p}{2} -b\right)!} \left( \frac{1-p}{2} \right)_{b+\frac{p}{2}}2b\cdot I(b) 
- \sum_{b=0}^{\frac{p}{2}} \frac{(-1)^{b+\frac{p}{2}}2^{2b}p!}{(2b)! \left( \frac{p}{2} -b\right)!} \left( \frac{1-p}{2} \right)_{b+\frac{p}{2}}(2b+1)\cdot I(b+1) \right].
\end{align}
Using the results for $J(b)$ and $I(b)$ from Appendix B, one obtains
\begin{align}
A^{(1)}= \frac{4\pi Q}{R\gamma} \cdot S^{(1)}_1, 
\end{align}
\begin{align}
C^{(1)} = -\frac{8\pi}{R} Q \gamma \beta \dot{\beta} \cdot( S^{(1)}_2 -
S^{(1)}_3),
\end{align}
where the series $S^{(1)}_i,$ $i=\overline{1,3}$ and their sums are written in Eqs. (C1) - (C3) of Appendix C.

Using Eqs. (29), (33) in Eq. (23), one obtains
\begin{align}
F_x^{(1)}(n=p)= &-\frac{Q^2 \ddot{w}}{2c^2R\gamma} \cdot s_1 -\frac{2Q^2 \ddot{w}}{c^2R\gamma} \cdot s_2 -\frac{2Q^2\ddot{w}}{c^2R\gamma} \cdot s_3+\frac{8Q^2 \ddot{w}}{3R c^2} \gamma \beta^4 \cdot s_4 +\frac{2Q^2\ddot{w}}{3Rc^2} \gamma \beta^4 \cdot s_5 \nonumber \\
&-\frac{4Q^2 \ddot{w}}{Rc^2} \gamma \beta^2 \cdot s_6 -\frac{Q^2 \ddot{w}}{Rc^2} \gamma \beta^2 \cdot s_7- \frac{8Q^2\ddot{w}}{5Rc^2} \gamma \beta^4 \cdot s_8 -\frac{2Q^2\ddot{w}}{5Rc^2} \gamma \beta^4 \cdot s_9,
\end{align}
where the series $s_i$, $i=\overline{1,9}$  and their sums are given in Appendix C.
Using the results from Eqs. (C4) - (C12)  in Eq. (31), one obtains
\begin{align}
F_x^{(1)}(n=p)=\frac{Q^2 \ddot{w}}{Rc^2\gamma} \left[ -\frac{(43 \beta^4-70\beta^2+75)}{30(1-\beta^2)^3}-\frac{\ln (1-\beta^2)}{(1-\beta^2)} +\frac{2\gamma}{\beta} \arcsin \beta\right].
\end{align}

\subsection*{III.C Calculation of $F_x^{(1)}(n=p+1)$}
Putting $n=p+1$ in Eq. (20), one obtains
\begin{align}
F_x^{(1)}(n=p+1)= \frac{Q}{4\pi} \sum_{p=0}^{\infty} \frac{(-1)^{p+2}}{(p+1)! \;p! \;c^{p+3}} \frac{p+2}{p+3} \int d {\bf r'} \rho({\bf r'},t) \frac{\partial^{p+2}}{\partial t ^{p+2}} \left[ w^p(t) \dot{w}(t) J_1(p,p+1,{\bf r'}) \right],
\end{align}
where
\begin{align}
J_1(p,p+1,{\bf r'})= \int_0^{\pi} d \theta \int_0^{2\pi} d \phi \;\sin \theta \; \partial_x^p |{\bf r}- {\bf r'}|^p.
\end{align}
We use in Eq. (33)
\begin{align}
\frac{\partial^{p+2}}{\partial t^{p+2}} \left[w^p(t) \dot{w}(t) J_1(p,p+1,{\bf r'})\right]=&\frac{(p+2)!}{2}\dot{w}^{p+1} \frac{\partial^2 J_1}{\partial t^2}+ (p+2)!\frac{(p+2)}{2} \dot{w}^p \ddot{w} \frac{\partial J_1}{\partial t}+ \nonumber \\
& \left[ \frac{(p+2)!(p+3)}{6}  \dot{w}^p \dot{\ddot{w}} +\frac{p(p+2)!(p+3)}{8}\dot{w}^{p-1} \ddot{w}^2  \right]J_1,
\end{align}
and note that only $p$ even gives non-zero contribution to Eq. (33). As
\begin{align}
J_1(p,p+1,{\bf r'}) \stackrel{\text{p even}}{=} \sum_{b=0}^{\frac{p}{2}} \frac{(-1)^{b+\frac{p}{2}} 2^{2b}p!}{(2b)! \left(\frac{p}{2}-b \right)!} \left( - \frac{p}{2}\right)_{b+\frac{p}{2}} \int_0^{\pi} d \theta \int_0^{2\pi} d \phi \sin \theta  \frac{(x-x')^{2b}}{|{\bf r}-{\bf r'}|^{2b}}_{|\eta=\eta_0},
\end{align}
we see that, because of the Pochhammer symbol $\left( - \frac{p}{2}\right)_{b+\frac{p}{2}}$, only $b=0$ gives non-zero contribution to the above equation, so
\begin{align}
J_1(p,p+1,{\bf r'}) \stackrel{\text{p even}}{=} 4\pi p!.
\end{align}

\noindent
Let us see if $\frac{\partial J_1}{\partial t}$ from Eq. (35) gives any contribution to $F_x^1(n=p+1)$. \\
For $p$ even, 
\begin{align}
\frac{\partial J_1(p, p+1, {\bf r'})}{\partial t}\stackrel{\text{p even}}{=} 2R\gamma \beta \dot{\beta} \sum_{b=0}^{\frac{p}{2}} \frac{(-1)^{b+\frac{p}{2}} 2^{2b} p! \;b}{(2b)! \left( \frac{p}{2}-b \right)!} \left( - \frac{p}{2} \right)_{b+\frac{p}{2}} \int_0^{\pi} d \theta\int_0^{2\pi} d \phi \sin \theta \cos \theta \nonumber \\
\cdot \left[ \frac{(x-x')^{2b-1}}{|{\bf r} - {\bf r'}|^{2b}} - \frac{(x-x')^{2b+1}}{|{\bf r} - {\bf r'}|^{2b+2}} \right]_{|\eta = \eta_0}.
\end{align}
The Pochhammer symbol $\left( - \frac{p}{2} \right)_{b+\frac{p}{2}}$, is different from zero only for $b=0$ and, because of the factor $b$ from the rhs of Eq. (38), it follows that 
\[\frac{\partial J_1(p, p+1, {\bf r'})}{\partial t}\stackrel{\text{p even}}{=}0.\]
For $p$ odd, 
\begin{align}
\frac{\partial J_1(p, p+1, {\bf r'})}{\partial t}\stackrel{\text{p odd}}{=} R\gamma \beta \dot{\beta} \sum_{b=0}^{\frac{p-1}{2}} \frac{(-1)^{b+\frac{p+1}{2}} 2^{2b+1} p! }{(2b)! \left( \frac{p-1}{2}-b \right)!} \left( - \frac{p}{2} \right)_{b+\frac{p}{2}} \int_0^{\pi} d \theta\int_0^{2\pi} d \phi \sin \theta \cos \theta \nonumber \\
\cdot \left[ \frac{(x-x')^{2b}}{|{\bf r} - {\bf r'}|^{2b+1}} - \frac{(x-x')^{2b+2}}{|{\bf r} - {\bf r'}|^{2b+3}} \right]_{|\eta = \eta_0}.
\end{align}
When we introduce Eq. (39) in Eqs. (35) and (33), we arrive at integrals of the form
\begin{align}
K= \int_0^{\pi} d \theta\int_0^{\pi} d \theta' \sin \theta \cos \theta \sin \theta' (\cos\theta- \cos \theta')^{2n} P_l(\cos \theta) P_l(\cos \theta'). 
\end{align}
Changing the integration variables in the above integral $\theta \rightarrow x$, $\theta' \rightarrow x'$ and then expanding $(x-x')^n$ using the binomial expansion formula, we obtain
\begin{align}
K=\sum_{k=0}^{2n} \frac{(2n)!}{k! (2n-k)!} \int_{-1}^1 dx \;x^{k+1} P_l(x) \int_{-1}^1 dx'\; x'^{2n-k} P_l(x').
\end{align}
As the fisrt integral in the rhs of the above equation is different from zero only when $k$ and $l$ have different parities, and the second integral in the rhs of the above equation is different from zero only when $k$ and $l$ have the same parity, it follows that $K=0$. We conclude that $\frac{\partial J_1}{\partial t}$ does not give any contribution to $F_x^1(n=p+1)$. Using a similar reasoning, it follows that neither $\frac{\partial^2 J_1}{\partial t^2}$  gives any contribution to $F_x^1(n=p+1)$.

Introducing Eq. (37) in Eqs. (35) and (33) and using \cite{wol}
\begin{align}
\sum_{p=0}^{\infty} \beta^{2p}(p+1)^2=\gamma^6 (1+\beta^2)
\end{align}
and
\begin{align}
\sum_{p=0}^{\infty}\beta^{2p} p (p+1)^2= 2 \gamma^8 \beta^2 (\beta^2+2),
\end{align}
one obtains
\begin{align}
F_x^{(1)}(n=p+1) = \frac{2}{3} \frac{Q^2 \dot{\ddot{w}}}{c^3}\gamma^6(1+\beta^2)+ \frac{2Q^2}{c^5} \ddot{w}^2 \dot{w} \gamma^8(\beta^2+2). 
\end{align}

\section{Calculation of $F_x^{(2)}(t)$}
The $x$ - component of Eq. (12) reads
\begin{align}
F_x^{(2)}(t)= \sum_{n=0}^{\infty} \frac{(-1)^{n}}{n!\,c^{n+2}} \frac{n-1}{n+2}
 \int d {\bf r} \int d {\bf r'}  \rho({\bf r'},t) \;| {\bf r}- {\bf r'}|^{n-3}(x-x')^2\frac{\partial^{n+1}}{\partial t^{n+1}} \left( \rho({\bf r},t) \dot{w}(t)\right). 
\end{align}
Using Eqs. (6), (13), the integral over ${\bf r}$ can be written as
\begin{align}
\int d {\bf r}\; 	|{\bf r} - {\bf r'}|^{n-3} (x-x')^2\frac{\partial^{n+1}}{\partial t^{n+1}} \left( \rho({\bf r},t) \dot{w}(t) \right) = \frac{Q}{4\pi} \sum_{p=0}^{\infty} \frac{1}{p!} \frac{\partial^{n+1}}{\partial t^{n+1}}\left[ w^p(t) \dot{w}(t) J_2(p,n;{\bf r'}) \right],
\end{align}
where
\begin{align}
J_2(p,n;{\bf r'})= \int_0^{\pi} d \theta \int_0^{2\pi} d \phi \sin \theta \; \partial_x^p \left[ | {\bf r}-{\bf r'}|^{n-3} (x-x')^2\right]_{|\eta=\eta_0} 
\end{align}
Just like $J_1(p,n,{\bf r'})$ from Eq. (19), $J_2(p,n,{\bf r'})$ depends on time, but we will omit its argument $t$ throughout this paper.
From Eqs. (45) - (47), it follows
\begin{align}
F_x^{(2)}(t)=\frac{Q}{4\pi} \sum_{p=0}^{\infty} \sum_{n=0}^{\infty} \frac{(-1)^n}{n! p! c^{n+2}} \frac{n-1}{n+2} \int d {\bf r'} \rho({\bf r'},t) \frac{\partial^{n+1}}{\partial t^{n+1}}\left[ w^p(t) \dot{w}(t) J_2(p,n;{\bf r'}) \right].
\end{align}

\subsection*{IV.A Calculation of $F_x^{(2)}(n=p-1)$}
From Eq. (47) it follows that
\begin{align}
J_2(p, p-1; {\bf r'})= \int_0^{\pi} d\theta \int_0^{2\pi} d \phi \sin \theta \left[ (x-x')^2 \partial_x^p |{\bf r} - {\bf r'}|^{p-4} +2p (x-x') \partial_x^{p-1} |{\bf r} - {\bf r'}|^{p-4} \right. \nonumber \\
\left. + p(p-1) \partial_x^{p-2} |{\bf r} - {\bf r'}|^{p-4}\right]_{|\eta= \eta_0}.
\end{align}
For $p$ odd, the series of $F_x^{(2)}$ that is obtained by putting $n=p-1$ in Eq. (48) contains integrals of the type 
\[ \int_0^{\pi} d \theta \int_0^{2\pi} d \phi \int_0^\pi d \theta' \int_0^{2\pi} d \phi'\sin \theta \sin \theta' (x-x')^{2k+1}=0, \]
and for $p$ even, all the derivatives in the rhs of Eq.(49) are zero. It follows that $F_x^{(2)}(n=p-1)=0.$

\subsection*{IV.B Calculation of $F_x^{(2)}(n=p)$}
Putting $n=p$ in Eq. (48) one obtains:
\begin{align}
F_x^{(2)}(n=p)=\frac{Q}{4\pi}\sum_{p=0}^{\infty} \frac{(-1)^p}{p! p! c^{p+2}} \frac{p-1}{p+2} \int d {\bf r'} \rho({\bf r'},t) \frac{\partial^{p+1}}{\partial t^{p+1}} \left[ w^p(t) \dot{w}(t) J_2(p, p, {\bf r'}) \right],
\end{align}
where
\begin{align}
J_2(p, p, {\bf r'})= \int_0^{\pi} d \theta \int_0^{2\pi} d \phi \; \sin \theta \; \partial_x^p \left[ | {\bf r}-{\bf r'}|^{p-3} (x-x')^2\right]_{|\eta=\eta_0}. 
\end{align}
Using
\begin{align}
\frac{\partial^{p+1}}{\partial t^{p+1}} \left[ w^p \dot{w} J_2\right]=(p+1)! \;\dot{w}^{p+1} \frac{\partial J_2}{\partial t} +(p+1)! \left( 1+\frac{p}{2} \right) \dot{w}^p \ddot{w} J_2
\end{align}
and noting that $p$ odd does not give any contribution, we obtain
\begin{align}
F_x^{(2)}(n=p) = \frac{Q \ddot{w}}{8 \pi c^2} \sum_{\substack{p=0 \\ p \;\text{even}}}^{\infty}\frac{(-1)^p \beta^p}{p!} (p^2-1) A^{(2)}  +\frac{Q \dot{w}}{4 \pi c^2} \sum_{\substack{p=0 \\ p \;\text{even}}}^{\infty} \frac{(-1)^p \beta^p}{p!} \frac{p^2-1}{p+2} C^2, 
\end{align}
where
\begin{align}
A^{(2)}=\int d {\bf r'} \rho({\bf r'},t)\, J_2(p, p, {\bf r'}),
\end{align}
\begin{align}
C^{(2)}=\int d {\bf r'} \rho({\bf r'},t)\, \frac{\partial J_2(p, p, {\bf r'})}{\partial t}.
\end{align}
Using in the expression of $J_2(p, p, {\bf r'})$ from Eq. (51)
\begin{align}
\partial_x^p \left[ | {\bf r}- {\bf r'}|^{p-3}(x-x')^2\right]=& (x-x')^2 \partial_x^p | {\bf r}- {\bf r'}|^{p-3} +2p (x-x') \partial_x^{p-1} | {\bf r}- {\bf r'}|^{p-3}\nonumber \\
& +p(p-1) \partial_x^{p-2} | {\bf r}- {\bf r'}|^{p-3}, 
\end{align}
$A^{(2)}$ and $C^{(2)}$ can be expressed in terms of the quantities $J(b)$ and $I(b)$ from Appendix B respectively as follows
\begin{align}
A^{(2)}=&\frac{Q}{4\pi} \left[ \sum_{b=0}^{\frac{p}{2}} \frac{(-1)^{b+\frac{p}{2}}2^{2b}p!}{(2b)! \left( \frac{p}{2}-b \right)!} \left( \frac{3-p}{2}\right)_{b+\frac{p}{2}} \cdot J(b+1) \right.\nonumber \\
& +4 \sum_{b=0}^{\frac{p}{2}-1} \frac{(-1)^{b+\frac{p}{2}}2^{2b}p!}{(2b+1)! \left( \frac{p}{2}-b-1 \right)!} \left( \frac{3-p}{2}\right)_{b+\frac{p}{2}} \cdot J(b+1)\nonumber \\
& \left.  +\sum_{b=0}^{\frac{p}{2}-1} \frac{(-1)^{b+\frac{p}{2}-1}2^{2b}p!}{(2b)! \left( \frac{p}{2}-b-1 \right)!} \left( \frac{3-p}{2}\right)_{b+\frac{p}{2}-1} \cdot J(b) \right], 
\end{align}
\begin{align}
C^{(2)}= &\frac{Q}{4\pi} R \gamma \beta \dot{\beta} \left\lbrace \sum_{b=0}^{\frac{p}{2}} \frac{(-1)^{b+\frac{p}{2}} 2^{2b} p!}{(2b)! \left( \frac{p}{2}-b \right)!} \left( \frac{3-p}{2} \right)_{b+\frac{p}{2}} \left[ (2b+2) I(b+1)-(2b+3)I(b+2) \right]\right.\nonumber \\
&+4\sum_{b=0}^{\frac{p}{2}-1} \frac{(-1)^{b+\frac{p}{2}} 2^{2b} p!}{(2b+1)! \left( \frac{p}{2}-b -1\right)!} \left( \frac{3-p}{2} \right)_{b+\frac{p}{2}} \left[ (2b+2) I(b+1)-(2b+3)I(b+2) \right]\nonumber \\
&\left. + \sum_{b=0}^{\frac{p}{2}-1} \frac{(-1)^{b+\frac{p}{2}-1} 2^{2b} p!}{(2b)! \left( \frac{p}{2}-b -1\right)!} \left( \frac{3-p}{2} \right)_{b+\frac{p}{2}-1} \left[ 2b\; I(b)-(2b+1)I(b+1) \right]\right\rbrace. 
\end{align}
 Using for $J(b)$ and $I(b)$ the results from Appendix B, we arrive at 
\begin{align}
A^2( \text{p even})= \frac{Q}{4\pi} \left\lbrace   \frac{16\pi^2}{R\gamma}   \frac{(-1)^{\frac{p}{2}}p!}{\Gamma \left( \frac{3}{2}- \frac{p}{2} \right)} \cdot S^{(2)}_1 + \frac{64 \pi^2}{R \gamma}  \frac{(-1)^{\frac{p}{2}}p!}{\Gamma \left( \frac{3}{2}-\frac{p}{2} \right)} \cdot S^{(2)}_2   + \frac{16\pi^2}{R\gamma}   \frac{(-1)^{\frac{p}{2}-1}p!}{\Gamma \left( \frac{3}{2}- \frac{p}{2} \right)} \cdot S^{(2)}_3 \right\rbrace,
\end{align}
and
\begin{align}
C^2(\text{p even})= - \frac{8\pi Q}{R} \gamma \beta \dot{\beta} \frac{(-1)^{\frac{p}{2}}p!}{\Gamma \left( \frac{3}{2}-\frac{p}{2}\right)}(S^{(2)}_4- S^{(2)}_5+4 S^{(2)}_6-4 S^{(2)}_7-S^{(2)}_8+S^{(2)}_9),
\end{align}
where the series $S^{(2)}_i,$ $i=\overline{1,9}$ and their sums are given in Eqs. (C13) - (C21) from Appendix C.
Introducing  the results from Eqs. (C13) - (C21) in Eqs. (59), (60) and then (53) - (55), we obtain
\begin{align}
F_x^{2}(n=p)=f_{21}+f_{22}, 
\end{align}
where\begin{align}
f_{21}=\frac{Q^2\ddot{w}}{6c^2R\gamma}(-s_{211}+4s_{212}) - \frac{Q^2\ddot{w}}{2c^2R\gamma}(-s_{213}+4s_{214}) + \frac{Q^2\ddot{w}}{c^2R\gamma}(-s_{215}+4s_{216})\nonumber \\
- \frac{Q^2\ddot{w}}{2c^2R\gamma}(-s_{217}+4s_{218}),
\end{align}
\begin{align}
f_{22}=-\frac{\pi Q^2}{Rc}\gamma \beta^2 \dot{\beta}\left[ \frac{3\beta^2}{8}s_{221}-\frac{45\beta^2}{8} s_{222} -\frac{9\beta^2}{16}s_{223}+\frac{105\beta^2}{16}s_{224}+\frac{9\beta^2}{4}s_{225}\right.\nonumber \\
\left. -\frac{1}{8} s_{226}+ \frac{15}{8} s_{227} -\frac{3}{2} s_{228}- s_{229}+\frac{1}{2} s_{2210}    \right],
\end{align}
and the series $s_{21i}$, $s_{21i}$, $i=\overline{1,8}$, $j=\overline{1,10}$ from Eqs. (62), (63) are given in Eqs. (C22) - (C39) from Appendix C. Replacing the sums of the series $s_{21i}$ and $s_{21i}$ in Eqs. (62) - (63), after simplifications one obtains
\begin{align}
f_{21}= - \frac{Q^2 \ddot{w}}{2c^2R \gamma}\frac{(5\beta^4-34\beta^2+5)}{15(1-\beta^2)^3},
\end{align}
\begin{align}
f_{22}= -\frac{Q^2}{Rc}\gamma \beta^2 \dot{\beta}\left[ \frac{2(10\beta^6-51\beta^4 +50\beta^2 -15)}{15\beta^4 (1-\beta^2)^2}+\frac{1}{\beta^2} \ln (1-\beta^2)\right. \nonumber \\
\left. +\frac{2(1-\beta^2)^{3/2}}{\beta^5} \arcsin \beta\right].
\end{align}
Introducing Eqs. (64), (65) in Eq. (61), one obtains
\begin{align}
F^{(2)}_x(n=p)= & - \frac{Q^2 \ddot{w}}{c^2 R\gamma}\left[  \frac{45\beta^6-238\beta^4+205\beta^2-60}{30\beta^2(1-\beta^2)^3} +\frac{1}{1-\beta^2} \ln (1-\beta^2) \right.\nonumber \\
&\left. + \frac{2\sqrt{1-\beta^2}}{\beta^3} \arcsin \beta  \right]. 
\end{align}

\subsection*{IV.C Calculation of $F_x^{(2)}(n=p+1)$}
Putting $n=p+1$ in Eqs. (48) and (47), one obtains
\begin{align}
F_x^{(2)}(n=p+1)=\frac{Q}{4\pi} \sum_{p=0}^{\infty} \frac{(-1)^{p+1}}{(p+1)!p! c^{p+3}} \frac{p}{p+3} \int d{\bf r'} \frac{\partial^{p+2}}{\partial t^{p+2}} \left[w^p(t) \dot{w}(t) J_2(p,p+1,{\bf r'})\right], (F2.1)
\end{align}
\begin{align}
J_2(p,p+1,{\bf r'}) = \int_0^{\pi} d\theta \int_0^{2\pi} d \phi \;\sin \theta \;\partial_x^p \left[ |{\bf r}- {\bf r'}|^{p-2}(x-x')^2\right]_{|\eta=\eta_0}.
\end{align}
We use Eq. (35) for $J_1 \rightarrow J_2$, and the expanded form of $J_2$
\begin{align}
J_2(p,p+1,{\bf r'})=& \int_0^\pi d \theta \int_0^{2\pi} d \phi \sin \theta \left[ (x-x')^2 \partial_x^p |{\bf r}-{\bf r'}|^{p-2} \right.\nonumber \\
& \left. +2p (x-x') \partial_x^{p-1} |{\bf r}-{\bf r'}|^{p-2} +p(p-1) \partial_x^{p-2} |{\bf r}- {\bf r'}|^{p-2}\right].
\end{align}
Reasonings similar to those in Sec. III.C lead to the conclusion that the first and second derivatives with respect to time of $J_2(p,p+1,{\bf r'})$ give no contribution to $F_x^2(n=p+1)$. The only non-zero contribution to $F_x^2(n=p+1)$ is that of 
\begin{align}
J_2(p,p+1,{\bf r'})_{|\text{p even}}=4 \pi p!.
\end{align}
Introducing Eq. (70) in (67), after using \cite{wol}
\begin{align}
\sum_{p=0}^{\infty} \beta^{2p} p(p+1)=\frac{2\beta^2}{(1-\beta^2)^3}
\end{align}
and
\begin{align}
\sum_{p=0}^{\infty} \beta^{2p} p^2(p+1)=\frac{2\beta^2(2\beta^2+1)}{(1-\beta^2)^4},
\end{align}
one obtains
\begin{align}
F_x^{(2)}(n=p+1)= - \frac{4}{3} \frac{Q^2 \dot{\ddot{w}}}{c^3} \gamma^6\beta^2 -\frac{2Q^2\ddot{w}^2}{c^4} \gamma^8 \beta (2\beta^2+1).
\end{align}

\section{The total electromagnetic self-force}
From Eqs. (32), (44), (66) and (73), we obtain the first two non-zero terms in the expansion of the total electromagnetic self-force acting in  the laboratory frame on a Lorentz-contractible spherical shell in rectilinear motion 
\begin{align}
F_x(t) = F_x^{(1)}(t) + F_x^{(2)}(t)= \frac{2Q^2\ddot{w}}{Rc^2 \gamma}\left[\frac{22\beta^4-55\beta^2+15}{15\beta^2(1-\beta^2)^2} +\frac{2\beta^2-1}{\beta^3\sqrt{1-\beta^2}} \arcsin \beta-\frac{\ln (1-\beta^2)}{1-\beta^2} \right]\nonumber \\
+\frac{2Q^2}{3c^3} \gamma^4 \dot{\ddot{w}}+\frac{2Q^2}{c^5} \gamma^6 \dot{w} \ddot{w}^2 + {\cal O}(R).
\end{align}
The term of order $R^0$ in the above equation is the space part of the relativistic Lorentz-Abraham-Dirac (LAD) force particularized for the case of rectilinear motion. The term of order $1/R$, as far as we know, is new in the literature and it generalizes to arbitrary velocities the well-known approximation for low-velocities $-(2Q^2 \ddot{w})/(3Rc^2)$.

\noindent
In the limit $\beta \rightarrow 0$ one obtains from the above equation
\begin{align}
F_x(t) \stackrel{\beta \rightarrow 0}{\longrightarrow} \frac{2Q^2\ddot{w}}{Rc^2}\left( - \frac{1}{3} - \frac{9\beta^2}{10} -\frac{387 \beta^4}{280} \right)+\frac{2Q^2 \ddot{w}}{c^4} (\beta+3\beta^3) +\frac{2Q^2\dot{\ddot{w}}}{3c^3}(1+2\beta^2+3\beta^4)*{\cal O}(\beta^5).
\end{align}

\section{Discussion and conclusions}
We have studied the electromagnetic self-force of a Lorentz-contractible spherical shell of radius $R$ in arbitrary rectilinear motion, in the laboratory frame of reference. We expanded the self-force in powers of its characteristic length $R$ and calculated the first two terms of this expansion. Our final result given in Eq.(74).

In \cite{yag}, the electromagnetic self-force of a Lorentz-contractible spherical shell was also calculated up to the order $R^0$. The author of \cite{yag} approached the problem  considering the shell to be made up of infinitesimal charged particles that act on each other. To ensure that these infinitesimal moving charges form an oblate ellipsoid in the laboratory frame of reference, he had to correct their accelerations so that they are dependent on their positions (Eq. (A8) of \cite{yag}). After noticing that the Lorentz transformation of the self-force of an extended body in its instantaneously proper frame does not necessarily have to coincide with the self-force in the laboratory frame, the author of \cite{yag} calculates the self-force directly in the laboratory frame in his Appendix B. His final result is given in its Eq. (B.41). By particularization for the case of rectilinear motion and using our notations and Gaussian units, Eq. (B.41) of \cite{yag} reads
\begin{align}
F_x^{\text{Yaghjian}}(t)=& - \frac{2Q^2}{3Rc^2} \frac{d}{dt} (\gamma \dot{w}) \nonumber \\
& + \frac{2Q^2}{3c^3} \gamma^4 \dot{\ddot{w}} + \frac{2Q^2}{c^4} \gamma^6 \beta \ddot{w}^2.
\end{align}
In the limit $\beta \rightarrow 0$, we have
\begin{align}
&F_x^{\text{Yaghjian}}(t) \stackrel{\beta \rightarrow 0}{=} - \frac{2Q^2\ddot{w}}{3Rc^2} \left(1+\frac{3\beta^2}{2} \right) \nonumber \\
&+\frac{2Q^2}{c^3}\left[ \dot{\ddot{w}}\left( \frac{1}{3}+\frac{2\beta^2}{3} \right)+\frac{\ddot{w}^2}{c} \left( \beta + 3\beta^3 \right)\right] + {\cal O}(\beta^4).
\end{align}
If we compare the above equations with our result (74), (75), we can see that, while the terms of order $R^0$ are identical, our term of order $1/R$ coincides with the term of order $1/R$ from \cite{yag} only for $\beta=0$. To understand where this difference comes from, we have to look more carefully at the expansions of the fields used in \cite{yag}. As the presentation of the calculations in \cite{yag} is very concise, we are showing in our Appendix D how the expansion of the field of a point charge was obtained in \cite{yag}. We can see that Eq. (A7) in \cite{yag} was obtained by keeping only the terms up to $1/c^4$ and $1/c^3$ from our equations (D1) and  (D2) respectively. As shown in our appendix D, the terms of higher orders from (D1), (D2) also give contributions of order $1/R$ and $R^0$ that depend on $\beta$ and its powers and, by truncating the series (D1), (D2), these contributions are lost. In fact, as noted in the last paragrapf from page 1051 of \cite{gor}, the expansions (D1), (D2) are good for low velocities, because they are expansions in powers of $v/c$ and, if we want some good expansions  for arbitrary velocities, we have to  regroup the terms. 

The expansions of the electromagnetic potentials and fields of a moving  point charge, that are valid for arbitrary velocities, were written in \cite{hni1}, \cite{hni2}. It can be easily verified that if in Eqs. (8), (9) of \cite{hni1} we use
\begin{align}
w^n \left( t - \frac{r}{c} \right) = \sum_{k=0}^{\infty} \frac{(-1)^k r^k}{c^k} \frac{d^k}{dt^k} w^n(t),
\end{align}
and note that
\begin{align}
w^n(t) \dot{w}(t)= \frac{1}{n+1} \frac{d w^{n+1}(t)}{dt},
\end{align}
after changing the order of summations and performing the sums over $n$, one obtains Eqs. (D1), (D2). We can say therefore that the expansions (8), (9) from \cite{hni1} are obtained from (D1), (D2) by regrouping the terms so that they become good for arbitrary velocities.

As noted in our Abstract and Sec. I, the method of calculation of the electromagnetic self-force used in this paper avoid some important difficulties of other approaches from the literature. We list the advantages of                                                                                                                                 our method in the following four points.\\
1) By working directly in the laboratory frame of reference, we avoid the Lorentz transformation of the total self-force acting on the shell from the instantaneous proper frame to the laboratory frame. As noted in \cite{yag}, for a shell of finite dimensions, the Lorentz transformation of the proper-frame self-force does not necessarily coincide with the self-force in the laboratory frame.\\
2) By using the expansion of the  the electromagnetic field in powers of $R$ instead of powers of $v/c$ as in other approaches, this method is appropriate for arbitrary high velocities.\\
3) By describing the shell by a time-dependent volume charge density instead of considering it as consisting of infinitesimal charged particles, we avoid all the approximations that are used when one "adjusts" the velocity and the acceleration of each constituent particle  so that, in the laboratory frame, the shell is Lorentz-contracted in the $x$-direction. Such formulas for adjusting the proper-frame accelerations of the constituent points of a rigid body can be found, for example, in \cite{yag}, \cite{lyl}, \cite {nik}, \cite{gro}. As emphasized in the last paragraph of Sec. 1.1 from \cite{lyl}, the adjusting of the accelerations of the constituent points of a Born-rigid body can be done consistently only in the approximation of very small dimensions. If we denote by $a_c=\ddot{w}(t)$ the acceleration of the shell's center and by $a$ the acceleration of a point on the shell that is located at a distance $x$ from the vertical axis of symmetry of the spheroid, the formula most used in the literature that connect these two accelerations is \cite{equ}
\begin{align}
a=\frac{a_c}{1+\frac{x a_c}{c^2}} \simeq a_c-x \frac{a_c^2}{c^2}.
\end{align}
Instead of using the above approximate formula, which is valid only for very small values of $R$, we used the charge density (6) that can be obtained by taking the Lorentz transformation of the charge density of a spherical shell from the instantaneous rest-frame to the laboratory frame. Let us see what are the velocity and the acceleration of a point on a shell in our approach. This is easily to calculate by taking the first and the second derivative with respect to time of $x$ from our Eq. (2) and remembering that $\gamma=1/\sqrt{1-\beta^2}=1/\sqrt{1-\dot{w}^2(t)/c^2}$ is dependent on time. Denoting by $v_c(t)= \dot{w}(t)$ the center's velocity, and $a_c(t)= \ddot{w}(t)$ the center's acceleration, one obtains
\begin{align}
v= v_c-x \gamma^2 \beta \dot{\beta},
\end{align}
\begin{align}
a = a_c-\frac{\gamma^2 x}{c^4}(c^2 a_c^2+ c^2 v_c \dot{a}_c+\gamma^2 v_c^2 a_c^2).
\end{align}
In the instantaneous proper frame we have $v_c=0$, and the above formula becomes
\begin{align}
a=a_c-x \frac{a_c^2}{c^2}.
\end{align}
This is the exact value of the acceleration of a point on the shell in the instantaneous frame of reference in our model.

Instead of considering the spherical shell as being composed of infinitesimal particles that interact with each other, as is usual in the literature, we considered it as being described by a time-dependent charge density.  Within this model we were able to write the electromagnetic self-force of a Lorentz-contractible spherical shell in rectilinear arbitrary motion as a series expansion in powers of $R$, and we calculated the first two non-zero terms of this expansion. The algorithm presented in this paper can be used for calculating the higher order terms, too.\\

\vspace{0.2cm}
\noindent
{\bf Acknowledgement:} The author is indebted to Professor Vladimir Hnizdo for his very useful observations regarding this work.

\appendix
\section{The expansion of the inverse distance between two points on an oblate spheroid raised to a power}
Let us consider two points on an oblate spheroid described by the position vectors ${\bf r}(x,y,z)$, ${\bf r'}(x',y',z')$. Using the parametrization given in Eq. (2), we can write
\begin{align}
\frac{1}{|{\bf r}-{\bf r'}|_{\text{sph}}^n}= \frac{1}{R^n2^{\frac{n}{2}}}\left[ 1-\cos \theta_{12}-\frac{\beta^2}{2}(\cos \theta - \cos \theta')^2 \right]^{-\frac{n}{2}}
\end{align}
where we used the notation 
\begin{align}
\cos \theta_{12}=\cos \theta \cos \theta'+2\sin \theta \sin \theta' \cos(\phi-\phi').
\end{align}
 Expanding (A1) in powers of $\beta$, one obtains
\begin{align}
\frac{1}{|{\bf r}-{\bf r'}|_{\text{sph}}^n}= \frac{1}{R^n2^{\frac{n}{2}}} \sum_{s=0}^{\infty}\frac{\beta^{2s}}{2^ss!} \frac{\Gamma\left(s+\frac{n}{2} \right)}{\Gamma\left( \frac{n}{2}\right)} \frac{( \cos \theta - \cos \theta')^{2s}}{(1-\cos \theta_{12})^{s+\frac{n}{2}}}.
\end{align}
But we proved in Eq. (69) of \cite{vam2} that
\begin{align}
 \frac{1}{(1-\cos \theta_{12})^{s+\frac{n}{2}}} = \sum_{k=0}^{\infty} \sum_{q=0}^{\left[ \frac{k}{2} \right]} \sum_{m=-k+2q}^{k-2q}\frac{\left( k+\frac{n}{2}+s-1\right)!}{\left( \frac{n}{2}+s-1\right)!} \nonumber \\
 \cdot \frac{(2k-4q+1)(k-2q-m)!}{2^q q! (2k-2q+1)!!(k-2q+m)!} P_{k-2q}^m(\cos \theta)\nonumber \\
 \cdot P_{k-2q}^m(\cos \theta')  e^{im(\phi - \phi')},
\end{align}
where $P_l^m$ are the associated Legendre functions \cite{bat}. Introducing (A4) in (A3) leads to Eq.(14) of this paper.

\section{Calculation of the integrals J(b) and I(b)}

\begin{align}
J(b)\equiv  \int_0^\pi d \theta \int_0^{2\pi} d \phi \int_0^\pi d \theta' \int_0^{2\pi} d \phi' \sin \theta \sin \theta' \frac{(x-x')^{2b}}{|{\bf r}- {\bf r'}|^{2b+1}_{|\eta=\eta'=\eta_0}}
\end{align}

\begin{align}
I(b)=\int_0^{\pi} d \theta \int_0^{2\pi} d \phi \int_0^{\pi} d \theta' \int_0^{2\pi} d\phi'  sin \theta \cos \theta \sin \theta' \frac{(x-x')^{2b-1}}{|{\bf r}- {\bf r'}|^{2b+1}}_{|\eta=\eta'=\eta_0}
\end{align}

\noindent
\underline{Calculation of $J(b)$}\\
Using Eq. (14) and performing the integrals over $\theta$ and $\phi$ as in \cite{vam2}, one obtains
\begin{align}
&J(b)=\frac{4\pi^2}{R \left(b-\frac{1}{2} \right)! \gamma^{2b} 2^{b+\frac{1}{2}}} \sum_{s=0}^{\infty} \sum_{k=0}^{\infty} \sum_{q=0}^{\left[ \frac{k}{2}\right]} \frac{\beta^{2s}}{2^s s!} \nonumber \\
&\cdot \frac{\left( k+s+b- \frac{1}{2}\right)!(2k-4q+1)(-1)^k 2^{2b+2s+2}}{2^q q! (2k-2q+1)!!(2b+2s+1)} \nonumber \\
&\cdot \frac{((b+s)!)^2}{(b+s-k+2q)!(b+s+k-2q+1)!}.
\end{align}
We change the summation order as follows
\begin{align}
\sum_{k=0}^{\infty} \sum_{q=0}^{\left[ \frac{k}{2}\right]} (\dots)= \sum_{q=0}^{\infty} \sum_{k=2q}^{\infty}(\dots),
\end{align}
and then change the summation index $k\rightarrow i$, $k-2q=i$. One obtains
\begin{align}
&J(b)=\frac{4\pi^2}{R\left( b-\frac{1}{2}\right)! \gamma^{2b}2^{b+\frac{1}{2}}} \sum_{s=0}^{\infty} \sum_{i=0}^{b+s} \sum_{q=0}^{\infty} \frac{\beta^{2s}}{2^s s!} \nonumber \\
&\cdot \frac{ \left( 2q+i+s+b- \frac{1}{2} \right)!(2i+1) (-1)^i 2^{2b+2s+2}}{2^q q! (2q+2i+1)!! (2b+2s+1)} \nonumber \\
&\cdot \frac{((b+s)!)^2}{(b+s-i)!(b+s+i+1)!}.
\end{align}
After writing the factorials in terms of Pochhammer functions and using the definition of the Gauss hypergeometric function, the summation over $q$ can be done immediately as follows
\begin{align}
\sum_{q=0}^{\infty}\frac{\left( 2q+i+s+b -\frac{1}{2} \right)!}{2^q q! (2q+2i+1)!!} = \frac{\left( i+s+b- \frac{1}{2}\right)! }{2^{s+b+\frac{1}{2}} } \nonumber \\
\cdot \frac{\left( -s-b- \frac{1}{2}\right)!}{\left( i-s-b+\frac{1}{2}\right)!}.
\end{align}
Then, the summation over $i$ can be done and the result can be written in terms of the generalized hypergeometric function \cite{pru}
\begin{align}
{}_4F_3 \left(  \begin{array}{llll}
-b-s, &1, & \frac{3}{2}, & s+b+\frac{1}{2}\\
 & \frac{1}{2}, & b+s+2, & \frac{3}{2}-s-b
\end{array} \bigg| 1\right) \nonumber\\
=(b+s+1)(1-2b-2s).
\end{align}
Finally, the summation over $s$ can be easily done in terms of the Gaussian hypergeometric function, and one obtains
\begin{align}
J(b)= \frac{16 \pi^2}{R \gamma (2b+1)} {}_2F_1 \left(  \begin{array}{cc}
1, &1\\
 & b+\frac{3}{2}
\end{array} \bigg| \beta^2\right).
\end{align}

\noindent
\underline{Calculation of $I(b)$}\\
After using in Eq. (B2) the expansion given in Eq. (14), the integrals over $\phi$ and $\phi'$ can be done as 
\begin{align}
 \int_0^{2\pi} d \phi \int_0^{2\pi} d \phi' e^{i m (\phi-\phi')}=16\pi^2 \delta_{m,0}. 
\end{align}
Changing the variables of integration $cos \theta \rightarrow x$, $cos \theta' \rightarrow y$, we have
\begin{align}
I(b)=-\frac{16 \pi^2}{R^2 \gamma^{2b-1}2^{b+\frac{1}{2}}} \sum_{s=0}^{\infty} \sum_{k=0}^{\infty} \sum_{q=0}^{\left[ \frac{k}{2}\right]} \frac{\beta^{2s} \Gamma \left(s+b+\frac{1}{2}\right)}{2^s s! \Gamma \left(b+\frac{1}{2}\right)} \frac{\left( k+s+b-\frac{1}{2}\right)! (2k-4q+1)}{\left(s+b-\frac{1}{2}\right)! 2^q q! (2k-2q+1)!!}\cdot D,
\end{align}
where
\begin{align}
D=\int_{-1}^1 dx \int_{-1}^1 dy \;x (x-y)^{2b+2s-1} P_{k-2q}(x) P_{k-2q}(y).
\end{align}
The double integral $D$ can be performed by a method similar to that indicated in \cite{vam1}, and one obtains
\begin{align}
D=\frac{(-1)^k 2 \sqrt{\pi} (b+s) (2b+2s-1)! (b+s)!}{\left( b+s+\frac{1}{2}\right)!(b+s-k+2q)!(k-2q+b+s+1)!}.
\end{align}
Then, the three remaining summations can be performed similarly to those from the calculation of $J(b)$, and one obtains
\begin{align}
I(b)=\frac{-32\pi^2}{R^2 (2b+1)} {}_2F_1 \left(  \begin{array}{cc}
1, &1\\
 & b+\frac{3}{2}
\end{array} \bigg| \beta^2\right). 
\end{align}

\section{Summation of some series}
\underline{ Series that appear in the calculation of $F_x^{(1)}(t)$}
\begin{align}
S^{(1)}_1=\sum_{b=0}^{\frac{p}{2}} \frac{(-1)^{b+\frac{p}{2}}2^{2b}p!}{(2b+1)! \left( \frac{p}{2}-b \right)!}\left(\frac{1-p}{2}\right)_{b+\frac{p}{2}} {}_2F_1 \left(  \begin{array}{cc}
1, &1\\
 & b+\frac{3}{2}
\end{array} \bigg| \beta^2\right) 
=\frac{\sqrt{\pi} \;p! \;(-1)^{\frac{p}{2}}}{\Gamma \left( \frac{1-p}{2}\right) \left( \frac{3}{2} \right)_{\frac{p}{2}}} {}_2F_1 \left(  \begin{array}{cc}
1, &1+\frac{p}{2}\\
 & \frac{3}{2}+\frac{p}{2}
\end{array} \bigg| \beta^2\right),
\end{align}
\begin{align}
S^{(1)}_2= &\sum_{b=0}^{\frac{p}{2}} \frac{(-1)^{b+\frac{p}{2}} 2^{2b} p!}{(2b)! \left( \frac{p}{2}-b \right)!} \left(\frac{1-p}{2}\right)_{b+\frac{p}{2}} \frac{2b}{2b+1} {}_2F_1 \left(  \begin{array}{cc}
1, &1\\
 & b+\frac{3}{2}
\end{array} \bigg| \beta^2\right)\nonumber \\
=&\frac{\pi \beta^2}{2} \frac{(-1)^{\frac{p}{2}} p!}{\Gamma \left( \frac{1-p}{2} \right) \Gamma \left( \frac{5}{2}+\frac{p}{2}\right)} {}_2F_1 \left(  \begin{array}{cc}
2, &\frac{p}{2}+1\\
 & \frac{5}{2}+\frac{p}{2}
\end{array} \bigg| \beta^2\right)
- \frac{\pi }{2} \frac{(-1)^{\frac{p}{2}} p!}{\Gamma \left( \frac{1-p}{2} \right) \Gamma \left( \frac{3}{2}+\frac{p}{2}\right)} {}_2F_1 \left(  \begin{array}{cc}
1, &\frac{p}{2}+1\\
 & \frac{3}{2}+\frac{p}{2}
\end{array} \bigg| \beta^2\right), 
\end{align}
\begin{align}
S^{(1)}_3= &\sum_{b=0}^{\frac{p}{2}} \frac{(-1)^{b+\frac{p}{2}} 2^{2b} p!}{(2b)! \left( \frac{p}{2}-b \right)!} \left(\frac{1-p}{2}\right)_{b+\frac{p}{2}} \frac{2b+1}{2b+3} \;{}_2F_1 \left(  \begin{array}{cc}
1, &1\\
 & b+\frac{5}{2}
\end{array} \bigg| \beta^2\right) \nonumber \\
=&\frac{3\pi \beta^2}{4} \frac{(-1)^{\frac{p}{2}} p!}{\Gamma \left( \frac{1-p}{2} \right) \Gamma \left( \frac{p}{2}+\frac{7}{2} \right)} {}_2F_1 \left(  \begin{array}{cc}
2, &\frac{p}{2}+1\\
 & \frac{p}{2}+\frac{7}{2}
\end{array} \bigg| \beta^2\right) 
-\frac{\pi}{2} \frac{(-1)^{\frac{p}{2}} p!}{\Gamma \left( \frac{1-p}{2}\right) \Gamma \left( \frac{p}{2}+ \frac{5}{2} \right)} {}_2F_1 \left(  \begin{array}{cc}
1, &\frac{p}{2}+1\\
 & \frac{p}{2}+\frac{5}{2}
\end{array} \bigg| \beta^2\right).
\end{align}
\begin{align}
s_1=\sum_{p=0}^\infty \beta^{2p} \frac{\left(\frac{1}{2}\right)_p}{\left(\frac{3}{2}\right)_p}\; {}_2F_1 \left(  \begin{array}{ll}
1, &p+1\\
 & p+\frac{3}{2}
\end{array} \bigg| \beta^2\right) = \frac{1}{1-\beta^2},
\end{align}
\begin{align}
s_2=\sum_{p=0}^\infty p \;\beta^{2p} \frac{\left(\frac{1}{2}\right)_p}{\left(\frac{3}{2}\right)_p}\; {}_2F_1 \left(  \begin{array}{ll}
1, &p+1\\
 & p+\frac{3}{2}
\end{array} \bigg| \beta^2\right) = \frac{\beta^2\gamma^4}{3},
\end{align}
\begin{align}
s_3= \sum_{p=0}^\infty p^2 \;\beta^{2p} \frac{\left(\frac{1}{2}\right)_p}{\left(\frac{3}{2}\right)_p}\; {}_2F_1 \left(  \begin{array}{ll}
1, &p+1\\
 & p+\frac{3}{2}
\end{array} \bigg| \beta^2\right) = \frac{\beta^2\gamma^6}{3} \left(1+\frac{\beta^2}{5}\right),
\end{align}
\begin{align}
s_4= \sum_{p=0}^\infty p \;\beta^{2p} \frac{\left(\frac{1}{2}\right)_p}{\left(\frac{5}{2}\right)_p}\; {}_2F_1 \left(  \begin{array}{ll}
2, &p+1\\
 & p+\frac{5}{2}
\end{array} \bigg| \beta^2\right) = \frac{\beta^2\gamma^4}{5},
\end{align}
\begin{align}
s_5=&\sum_{p=0}^\infty \frac{1}{p+1} \;\beta^{2p} \frac{\left(\frac{1}{2}\right)_p}{\left(\frac{5}{2}\right)_p}\; {}_2F_1 \left(  \begin{array}{ll}
2, &p+1\\
 & p+\frac{5}{2}
\end{array} \bigg| \beta^2\right) = \gamma^4+3\gamma^2+\frac{\gamma^4}{\beta^2} (6\beta^4-10\beta^2+3) \nonumber \\
&+\frac{3}{\beta^2}\; \ln(1-\beta^2)-\frac{3\gamma(1-2\beta^2)}{\beta^3}\;  \arcsin \beta,
\end{align}
\begin{align}
s_6= \sum_{p=0}^\infty p \;\beta^{2p} \frac{\left(\frac{1}{2}\right)_p}{\left(\frac{3}{2}\right)_p}\; {}_2F_1 \left(  \begin{array}{ll}
1, &p+1\\
 & p+\frac{3}{2}
\end{array} \bigg| \beta^2\right) = \frac{\beta^2\gamma^4}{3},
\end{align}
\begin{align}
s_7= \sum_{p=0}^\infty \frac{1}{p+1} \;\beta^{2p} \frac{\left(\frac{1}{2}\right)_p}{\left(\frac{3}{2}\right)_p}\; {}_2F_1 \left(  \begin{array}{ll}
1, &p+1\\
 & p+\frac{3}{2}
\end{array} \bigg| \beta^2\right) = \frac{1}{\beta^2} \ln(1-\beta^2)+\frac{2\gamma}{\beta} \arcsin \beta,
\end{align}
\begin{align}
s_8= \sum_{p=0}^\infty p \;\beta^{2p} \frac{\left(\frac{1}{2}\right)_p}{\left(\frac{7}{2}\right)_p}\; {}_2F_1 \left(  \begin{array}{ll}
2, &p+1\\
 & p+\frac{7}{2}
\end{array} \bigg| \beta^2\right) = \frac{(-2\beta^4-10\beta^2+15)}{6\beta^4(1-\beta^2)} - \frac{15}{6\beta^5} \arctanh \beta,
\end{align}
\begin{align}
s_9= \sum_{p=0}^\infty \frac{1}{p+1} \;\beta^{2p} \frac{\left(\frac{1}{2}\right)_p}{\left(\frac{7}{2}\right)_p}\; {}_2F_1 \left(  \begin{array}{ll}
2, &p+1\\
 & p+\frac{7}{2}
\end{array} \bigg| \beta^2\right) = &\frac{5}{\beta^2} \ln (1-\beta^2)-5\frac{(2\beta^2+1)}{\beta^5 \gamma} \arcsin \beta \nonumber \\
& +\frac{10}{\beta^5} \arctanh \beta - 5\frac{(1-\beta^2)}{\beta^4}.
\end{align}

\underline{ Series that appear in the calculation of $F_x^{(2)}(t)$}
\begin{align}
S^{(2)}_1=&\sum_{b=0}^{\frac{p}{2}} \frac{(-1)^b\; 2^{2b}\; \Gamma \left( \frac{3}{2}+b\right)}{(2b)! \left( \frac{p}{2}-b \right)! (2b+3)}\; {}_2F_1 \left(  \begin{array}{cc}
1, & 1\\
 &b+\frac{5}{2}
\end{array} \bigg| \beta^2\right) \nonumber \\
&=\frac{\pi}{8 \Gamma \left( \frac{p}{2}+\frac{5}{2}\right)} {}_2F_1 \left(  \begin{array}{cc}
1, & \frac{p}{2}+1\\
 &\frac{p}{2}+\frac{5}{2}
\end{array} \bigg| \beta^2\right) -\frac{3\pi}{8 \Gamma \left( \frac{p}{2}+\frac{5}{2}\right)} {}_2F_1 \left(  \begin{array}{cc}
1, & \frac{p}{2}\\
 &\frac{p}{2}+\frac{5}{2}
\end{array} \bigg| \beta^2\right),
\end{align}
\begin{align}
S^{(2)}_2=\sum_{b=0}^{\frac{p}{2}-1} \frac{(-1)^b\; 2^{2b}\; \Gamma \left( \frac{3}{2}+b\right)}{(2b+1)! \left( \frac{p}{2}-b-1 \right)! (2b+3)}\; {}_2F_1 \left(  \begin{array}{cc}
1, & 1\\
 &b+\frac{5}{2}
\end{array} \bigg| \beta^2\right) = \frac{\pi}{8\; \Gamma \left( \frac{3}{2}+\frac{p}{2} \right)} \;{}_2F_1 \left(  \begin{array}{cc}
1, & \frac{p}{2}\\
 & \frac{p}{2} + \frac{3}{2}
\end{array} \bigg| \beta^2\right),  
\end{align}
\begin{align}
S^{(2)}_3=\sum_{b=0}^{\frac{p}{2}-1} \frac{(-1)^b\; 2^{2b}\; \Gamma \left( \frac{1}{2}+b\right)}{(2b+1)! \left( \frac{p}{2}-b-1 \right)! }\; {}_2F_1 \left(  \begin{array}{cc}
1, & 1\\
 &b+\frac{3}{2}
\end{array} \bigg| \beta^2\right)  = \frac{\pi}{2\; \Gamma \left( \frac{1}{2}+\frac{p}{2} \right)} \;{}_2F_1 \left(  \begin{array}{cc}
1, & \frac{p}{2}\\
 & \frac{p}{2} + \frac{1}{2}
\end{array} \bigg| \beta^2\right),  
\end{align}
\begin{align}
S^{(2)}_4=\sum_{b=0}^{\frac{p}{2}} \frac{(-1)^b 2^{2b}}{(2b)! \left( \frac{p}{2}-b \right)!} \Gamma \left( \frac{3}{2}+b\right) \frac{2b+2}{2b+3}\; {}_2F_1 \left(  \begin{array}{cc}
1, & 1\\
 & b + \frac{5}{2}
\end{array} \bigg| \beta^2\right)\nonumber \\
=\frac{3\pi \beta^2}{8 \Gamma \left( \frac{7}{2}+\frac{p}{2} \right)}\;{}_2F_1 \left(  \begin{array}{cc}
2, & 1+\frac{p}{2}\\
 & \frac{p}{2} + \frac{7}{2}
\end{array} \bigg| \beta^2\right)- \frac{15 \pi \beta^2}{8 \Gamma \left( \frac{p}{2}+\frac{7}{2}\right)}{}_2F_1 \left(  \begin{array}{cc}
2, & \frac{p}{2}\\
 & \frac{p}{2} + \frac{5}{2}
\end{array} \bigg| \beta^2\right)\nonumber \\
- \frac{\pi }{8 \Gamma \left( \frac{5}{2}+\frac{p}{2} \right)}\;{}_2F_1 \left(  \begin{array}{cc}
1, & 1+\frac{p}{2}\\
 & \frac{p}{2} + \frac{5}{2}
\end{array} \bigg| \beta^2\right)+ \frac{3 \pi }{8 \Gamma \left( \frac{p}{2}+\frac{5}{2}\right)}{}_2F_1 \left(  \begin{array}{cc}
1, & \frac{p}{2}\\
 & \frac{p}{2} + \frac{5}{2}
\end{array} \bigg| \beta^2\right),
\end{align}
\begin{align}
S^{(2)}_5=\sum_{b=0}^{\frac{p}{2}} \frac{(-1)^b 2^{2b}}{(2b)! \left( \frac{p}{2}-b \right)!} \Gamma \left( \frac{3}{2}+b\right) \frac{2b+3}{2b+5}\; {}_2F_1 \left(  \begin{array}{cc}
1, & 1\\
 & b + \frac{7}{2}
\end{array} \bigg| \beta^2\right) \nonumber\\
= \frac{9 \pi \beta^2}{16 \Gamma \left( \frac{p}{2}+\frac{9}{2}\right)}{}_2F_1 \left(  \begin{array}{cc}
2, & \frac{p}{2}+1\\
 & \frac{p}{2} + \frac{9}{2}
\end{array} \bigg| \beta^2\right) + \frac{3 \pi }{2 \Gamma \left( \frac{p}{2}+\frac{7}{2}\right)}{}_2F_1 \left(  \begin{array}{cc}
1, & \frac{p}{2}\\
 & \frac{p}{2} + \frac{7}{2}
\end{array} \bigg| \beta^2\right) \nonumber \\
- \frac{105 \pi \beta^2}{16 \Gamma \left( \frac{p}{2}+\frac{9}{2}\right)}{}_2F_1 \left(  \begin{array}{cc}
2, & \frac{p}{2}\\
 & \frac{p}{2} + \frac{9}{2}
\end{array} \bigg| \beta^2\right), 
\end{align}
\begin{align}
S^{(2)}_6=\sum_{b=0}^{\frac{p}{2}-1} \frac{(-1)^b 2^{2b}}{(2b+1)! \left( \frac{p}{2}-b-1 \right)!} \Gamma \left( \frac{3}{2}+b\right) \frac{2b+2}{2b+3}\; {}_2F_1 \left(  \begin{array}{cc}
1, & 1\\
 & b + \frac{5}{2}
\end{array} \bigg| \beta^2\right)\nonumber \\
=\frac{3 \pi \beta^2}{8 \Gamma \left( \frac{p}{2}+\frac{5}{2}\right)}{}_2F_1 \left(  \begin{array}{cc}
2, & \frac{p}{2}\\
 & \frac{p}{2} + \frac{5}{2}
\end{array} \bigg| \beta^2\right)- \frac{ \pi }{8 \Gamma \left( \frac{p}{2}+\frac{3}{2}\right)}{}_2F_1 \left(  \begin{array}{cc}
1, & \frac{p}{2}\\
 & \frac{p}{2} + \frac{3}{2}
\end{array} \bigg| \beta^2\right) 
\end{align}
\begin{align}
S^{(2)}_7=\sum_{b=0}^{\frac{p}{2}-1} \frac{(-1)^b 2^{2b}}{(2b+1)! \left( \frac{p}{2}-b-1 \right)!} \Gamma \left( \frac{3}{2}+b\right) \frac{2b+3}{2b+5}\; {}_2F_1 \left(  \begin{array}{cc}
1, & 1\\
 & b + \frac{7}{2}
\end{array} \bigg| \beta^2\right)\nonumber \\
=\frac{15 \pi \beta^2}{16 \Gamma \left( \frac{p}{2}+\frac{7}{2}\right)}{}_2F_1 \left(  \begin{array}{cc}
2, & \frac{p}{2}\\
 & \frac{p}{2} + \frac{7}{2}
\end{array} \bigg| \beta^2\right)- \frac{ 3\pi }{8 \Gamma \left( \frac{p}{2}+\frac{5}{2}\right)}{}_2F_1 \left(  \begin{array}{cc}
1, & \frac{p}{2}\\
 & \frac{p}{2} + \frac{5}{2}
\end{array} \bigg| \beta^2\right) 
\end{align}
\begin{align}
S^{(2)}_8=\sum_{b=0}^{\frac{p}{2}-1} \frac{(-1)^b 2^{2b}}{(2b)! \left( \frac{p}{2}-b-1 \right)!} \Gamma \left( \frac{1}{2}+b\right) \frac{2b}{2b+1}\; {}_2F_1 \left(  \begin{array}{cc}
1, & 1\\
 & b + \frac{3}{2}
\end{array} \bigg| \beta^2\right)\nonumber \\
=- \frac{ \pi }{2 \Gamma \left( \frac{p}{2}+\frac{1}{2}\right)}{}_2F_1 \left(  \begin{array}{cc}
1, & \frac{p}{2}-1\\
 & \frac{p}{2} + \frac{1}{2}
\end{array} \bigg| \beta^2\right) 
\end{align}
\begin{align}
S^{(2)}_9=\sum_{b=0}^{\frac{p}{2}-1} \frac{(-1)^b 2^{2b}}{(2b)! \left( \frac{p}{2}-b-1 \right)!} \Gamma \left( \frac{1}{2}+b\right) \frac{2b+1}{2b+3}\; {}_2F_1 \left(  \begin{array}{cc}
1, & 1\\
 & b + \frac{5}{2}
\end{array} \bigg| \beta^2\right)\nonumber \\
=\frac{3 \pi \beta^2}{4 \Gamma \left( \frac{p}{2}+\frac{5}{2}\right)}{}_2F_1 \left(  \begin{array}{cc}
2, & \frac{p}{2}\\
 & \frac{p}{2} + \frac{5}{2}
\end{array} \bigg| \beta^2\right)- \frac{ \pi }{2 \Gamma \left( \frac{p}{2}+\frac{3}{2}\right)}{}_2F_1 \left(  \begin{array}{cc}
1, & \frac{p}{2}\\
 & \frac{p}{2} + \frac{3}{2}
\end{array} \bigg| \beta^2\right).
\end{align}
\begin{align}
s_{211}= \sum_{p=0}^{\infty} \beta^{2p} \frac{\left( -\frac{1}{2} \right)_p}{\left( \frac{5}{2} \right)_p} {}_2F_1 \left(  \begin{array}{cc}
1, & p+1\\
 &p+\frac{5}{2}
\end{array} \bigg| \beta^2\right)= \frac{3}{2\beta^3} \left[\beta-(1-\beta^2)\; \text{arctanh}\;\beta\right]
\end{align}
\begin{align}
s_{212}= \sum_{p=0}^{\infty} p^2\beta^{2p} \frac{\left( -\frac{1}{2} \right)_p}{\left( \frac{5}{2} \right)_p} {}_2F_1 \left(  \begin{array}{cc}
1, & p+1\\
 &p+\frac{5}{2}
\end{array} \bigg| \beta^2\right)= \frac{3-5\beta^2}{8\beta^2(1-\beta^2)}-\frac{3(1-\beta^2)}{8\beta^3} \; \text{arctanh}\; \beta
\end{align}
\begin{align}
s_{213}=\sum_{p=0}^{\infty} \beta^{2p} \frac{\left( -\frac{1}{2} \right)_p}{\left( \frac{5}{2} \right)_p}\; {}_2F_1 \left(  \begin{array}{cc}
1, & p\\
 &p+\frac{5}{2}
\end{array} \bigg| \beta^2\right)=\frac{4\beta^2-3}{2\beta^2}+\frac{3}{2}\frac{1-\beta^2}{\beta^3} \;\text{arctanh} \;\beta
\end{align}
\begin{align}
s_{214}= \sum_{p=0}^{\infty}p^2\; \beta^{2p} \frac{\left( -\frac{1}{2} \right)_p}{\left( \frac{5}{2} \right)_p}\; {}_2F_1 \left(  \begin{array}{cc}
1, & p\\
 &p+\frac{5}{2}
\end{array} \bigg| \beta^2\right)= \frac{-16\beta^4+25\beta^2-15}{40\beta^2(1-\beta^2)}\nonumber \\
+\frac{3}{8} \frac{(1-\beta^2)}{\beta^3}\; \text{arctanh} \;\beta
\end{align}
\begin{align}
s_{215}= \sum_{p=0}^{\infty} \beta^{2p} \frac{\left( -\frac{1}{2} \right)_p}{\left( \frac{3}{2} \right)_p}\; {}_2F_1 \left(  \begin{array}{cc}
1, & p\\
 &p+\frac{3}{2}
\end{array} \bigg| \beta^2\right)=2-\frac{1}{\beta}\; \text{arctanh}\; \beta
\end{align}
\begin{align}
s_{216}= \sum_{p=0}^{\infty} p^2\;\beta^{2p} \frac{\left( -\frac{1}{2} \right)_p}{\left( \frac{3}{2} \right)_p}\; {}_2F_1 \left(  \begin{array}{cc}
1, & p\\
 &p+\frac{3}{2}
\end{array} \bigg| \beta^2\right)= \frac{8\beta^4-15\beta^2+5}{20(1-\beta^2)^2}-\frac{1}{4\beta} \; \text{arctanh}\; \beta
\end{align}
\begin{align}
s_{217}= \sum_{p=0}^{\infty} \beta^{2p} \frac{\left( -\frac{1}{2} \right)_p}{\left( \frac{1}{2} \right)_p}\; {}_2F_1 \left(  \begin{array}{cc}
1, & p\\
 &p+\frac{1}{2}
\end{array} \bigg| \beta^2\right) = \frac{1-2\beta^2}{1-\beta^2}
\end{align}
\begin{align}
s_{218}= \sum_{p=0}^{\infty} p^2\;\beta^{2p} \frac{\left( -\frac{1}{2} \right)_p}{\left( \frac{1}{2} \right)_p}\; {}_2F_1 \left(  \begin{array}{cc}
1, & p\\
 &p+\frac{1}{2}
\end{array} \bigg| \beta^2\right) = -\frac{\beta^2(2\beta^4-5\beta^2+5)}{5(1-\beta^2)^3} 
\end{align}
\begin{align}
s_{221}= \sum_{p=0}^{\infty} (-1)^p \beta^{2p}\;\frac{4p^2-1}{p+1} \frac{1}{\Gamma \left( \frac{3}{2}-p\right) \Gamma \left( p+\frac{7}{2}\right)} {}_2F_1 \left(  \begin{array}{cc}
2, & p+1\\
 &p+\frac{7}{2}
\end{array} \bigg| \beta^2\right) \nonumber \\
=\frac{16}{3\pi} \left[ \frac{5\beta^2+3}{3\beta^4} +\frac{1}{\beta^2} \ln (1-\beta^2)- \frac{\sqrt{1-\beta^2}}{\beta^5} (2\beta^2+1) \arcsin \beta \right]
\end{align}
\begin{align}
s_{222}&= \sum_{p=0}^{\infty} (-1)^p \beta^{2p}\;\frac{4p^2-1}{p+1} \frac{1}{\Gamma \left( \frac{3}{2}-p\right) \Gamma \left( p+\frac{7}{2}\right)} {}_2F_1 \left(  \begin{array}{cc}
2, & p\\
 &p+\frac{7}{2}
\end{array} \bigg| \beta^2\right) \nonumber \\
&=- \frac{32(47\beta^4-40\beta^2-15)}{225\pi \beta^4}-\frac{16(4\beta^2-5)}{15\pi \beta^2} \ln (1-\beta^2)\nonumber \\ &-\frac{32(4\beta^2+1)(1-\beta^2)^{3/2}}{15\pi \beta^5} \arcsin \beta 
\end{align}
\begin{align}
s_{223}&= \sum_{p=0}^{\infty} (-1)^p \beta^{2p}\;\frac{4p^2-1}{p+1} \frac{1}{\Gamma \left( \frac{3}{2}-p\right) \Gamma \left( p+\frac{9}{2}\right)} {}_2F_1 \left(  \begin{array}{cc}
2, & p+1\\
 &p+\frac{9}{2}
\end{array} \bigg| \beta^2\right) \nonumber \\
&= \frac{32}{45 \pi \beta^6} (5\beta^4+4\beta^2-15) +\frac{64}{15\pi \beta^7} \;\text{arctanh} \; \beta +\frac{32}{15 \pi \beta^2} \ln (1-\beta^2)\nonumber \\
&+\frac{32}{15 \pi \beta^7} (1-\beta^2)^{3/2}(2\beta^2+3) \arcsin \beta 
\end{align}
\begin{align}
s_{224}&= \sum_{p=0}^{\infty} (-1)^p \beta^{2p}\;\frac{4p^2-1}{p+1} \frac{1}{\Gamma \left( \frac{3}{2}-p\right) \Gamma \left( p+\frac{9}{2}\right)} {}_2F_1 \left(  \begin{array}{cc}
2, & p\\
 &p+\frac{9}{2}
\end{array} \bigg| \beta^2\right) \nonumber \\
&=-\frac{64}{1575 \pi \beta^6} (47 \beta^6-62 \beta^4-30\beta^2+90)+ \frac{64}{35\pi \beta^7}\; \text{arctanh}\; \beta\nonumber \\
&+\frac{32}{105\pi \beta^2} (7-4\beta^2) \ln (1-\beta^2) +\frac{64}{105 \pi \beta^7}\frac{ (1-\beta^2)^3 (4\beta^2+3)}{\sqrt{1-\beta^2}} \arcsin \beta
\end{align}
\begin{align}
s_{225}&= \sum_{p=0}^{\infty} (-1)^p \beta^{2p}\;\frac{4p^2-1}{p+1} \frac{1}{\Gamma \left( \frac{3}{2}-p\right) \Gamma \left( p+\frac{5}{2}\right)} {}_2F_1 \left(  \begin{array}{cc}
2, & p\\
 &p+\frac{5}{2}
\end{array} \bigg| \beta^2\right) \nonumber \\
&=\frac{16(47\beta^4-65\beta^2+15)}{45\pi \beta^2 (1-\beta^2)} -\frac{8}{3\pi \beta^2} (4\beta^2-3) \ln (1-\beta^2)\nonumber \\
&-\frac{16}{3\pi \beta^3}(1-4\beta^2) \sqrt{1-\beta^2} \arcsin \beta
\end{align}
\begin{align}
s_{226}&= \sum_{p=0}^{\infty} (-1)^p \beta^{2p}\;\frac{4p^2-1}{p+1} \frac{1}{\Gamma \left( \frac{3}{2}-p\right) \Gamma \left( p+\frac{5}{2}\right)} {}_2F_1 \left(  \begin{array}{cc}
1, & p+1\\
 &p+\frac{5}{2}
\end{array} \bigg| \beta^2\right) \nonumber \\
&=\frac{16}{\pi \beta^2}+\frac{8}{\pi \beta^2} \ln (1-\beta^2) -\frac{16}{\pi \beta^3}\sqrt{1-\beta^2} \arcsin \beta
\end{align}
\begin{align}
s_{227}&= \sum_{p=0}^{\infty} (-1)^p \beta^{2p}\;\frac{4p^2-1}{p+1} \frac{1}{\Gamma \left( \frac{3}{2}-p\right) \Gamma \left( p+\frac{5}{2}\right)} {}_2F_1 \left(  \begin{array}{cc}
1, & p\\
 &p+\frac{5}{2}
\end{array} \bigg| \beta^2\right) \nonumber \\
&= -\frac{16(5\beta^2-6)}{9\pi \beta^2}-\frac{8}{3\pi \beta^2}(2\beta^2-3) \ln (1-\beta^2)-\frac{32}{3\pi \beta^3}(1-\beta^2)^{3/2} \arcsin \beta
\end{align}
\begin{align}
s_{228}&= \sum_{p=0}^{\infty} (-1)^p \beta^{2p}\;\frac{4p^2-1}{p+1} \frac{1}{\Gamma \left( \frac{3}{2}-p\right) \Gamma \left( p+\frac{7}{2}\right)} {}_2F_1 \left(  \begin{array}{cc}
1, & p\\
 &p+\frac{7}{2}
\end{array} \bigg| \beta^2\right) \nonumber \\
&= -\frac{32}{45 \pi \beta^4} (5\beta^4-11\beta^2+15) +\frac{32}{5 \pi \beta^5}\; \text{arctanh}\; \beta \nonumber \\
&+\frac{16}{15 \pi \beta^2}(5-2\beta^2) \ln (1-\beta^2)+ \frac{64}{15\pi \beta^5}(1-\beta^2)^{5/2} \arcsin \beta
\end{align}
\begin{align}
s_{229}&= \sum_{p=0}^{\infty} (-1)^p \beta^{2p}\;\frac{4p^2-1}{p+1} \frac{1}{\Gamma \left( \frac{3}{2}-p\right) \Gamma \left( p+\frac{3}{2}\right)} {}_2F_1 \left(  \begin{array}{cc}
1, & p\\
 &p+\frac{3}{2}
\end{array} \bigg| \beta^2\right) \nonumber \\
&=-\frac{8}{3\pi}\frac{(6-5\beta^2)}{(1-\beta^2)}-\frac{4}{\pi \beta^2} (2\beta^2-1) \ln (1-\beta^2)+\frac{16}{\pi \beta} \sqrt{1-\beta^2} \arcsin \beta
\end{align}
\begin{align}
s_{2210}&= \sum_{p=0}^{\infty} (-1)^p \beta^{2p}\;\frac{4p^2-1}{p+1} \frac{1}{\Gamma \left( \frac{3}{2}-p\right) \Gamma \left( p+\frac{1}{2}\right)} {}_2F_1 \left(  \begin{array}{cc}
1, & p-1\\
 &p+\frac{1}{2}
\end{array} \bigg| \beta^2\right) \nonumber \\
&= \frac{4}{15 \pi} \frac{(94 \beta^6-205 \beta^4+120 \beta^2-15)}{(1-\beta^2)^2}- \frac{32}{\pi}\beta \sqrt{1-\beta^2} \arcsin \beta\nonumber \\
& +\frac{2}{\pi} \frac{(8\beta^4-4\beta^2-1)}{\beta^2} \ln (1-\beta^2) 
\end{align}

\section{About Eq. (A.7) from \cite{yag}}
We consider in this section the general case of a point charged particle of charge $e$ moving along an arbitrary trajectory ${\bf w}(t)$.
Using the Lagrange expansion for retarded functions, its electromagnetic  potentials (Lienard - Wiechert potentials) can be written as follows (\cite{gor}, \cite{pag})
\begin{align}
\phi({\bf r},t)= e \sum_{k=0}^{\infty} \frac{(-1)^k}{c^k k!} \frac{d^k}{dt^k} R^{k-1},
\end{align}
\begin{align}
{\bf A}({\bf r},t)= \frac{e}{c} \sum_{k=0}^{\infty} \frac{(-1)^k}{c^k k!} \frac{d^k}{dt^k} \left( R^{k-1} {\bf v}\right),
\end{align}
where ${\bf R}= {\bf r} -{\bf w}(t)$, $R= |{\bf r} -{\bf w}(t)|$, ${\bf v}(t) =d {\bf w}(t)/dt$.
Using $d{\bf R}/dt=-{\bf v}(t)$, $dR/dt=-{\bf R} \cdot {\bf v}/R$,
one obtains the first terms of the above expansions
\begin{align}
\phi({\bf r},t)=e \left\lbrace \frac{1}{R}+ \frac{1}{2c^2} \left[ - \frac{({\bf R} \cdot {\bf v})^2}{R^3} - \frac{v^2}{R} - \frac{{\bf R}\cdot \dot{{\bf v}}}{R}\right] \right.\nonumber \\
  \left. + {\cal O}\left( \frac{1}{c^3}\right) \right\rbrace,
\end{align}
\begin{align}
{\bf A}({\bf r},t)= \frac{e}{c} \left[ \frac{{\bf v}}{R} -\frac{1}{c} \frac{d}{dt} {\bf v} + \frac{1}{2c^2} \frac{d^2}{dt^2} (R {\bf v})+ {\cal O}\left( \frac{1}{c^3} \right)\right].
\end{align}
Using
\begin{align}
{\bf E}({\bf r},t)= - \nabla \phi ({\bf r},t) -\frac{1}{c} \frac{\partial {\bf A}({\bf r},t)}{\partial t},
\end{align}
one obtains from (D3), (D4) for ${\bf v}=0$ the result
\begin{align}
{\bf E} ({\bf r},t)= e \left\lbrace \frac{R_x}{R^3} +\frac{1}{2Rc^2} \left[ \dot{\bf v}-\frac{{\bf R}({\bf R} \cdot \dot{{\bf v}})}{R^2} \right]+ {\cal O}\left( \frac{1}{c^3} \right) \right\rbrace.
\end{align}
This is Eq.(A.7) from \cite{yag}, where, for shortness, we have written only the terms of order $1/R^2$ and $1/R$. So, Eq. (A.7) from \cite{yag} was obtained by keeping the terms up to $1/c^4$ and $1/c^3$ in (D.1) and (D.2) respectively. But the higher order terms in (D1), (D2) also give contributions of order $1/R^2$ and $1/R$, that depend on $\beta$ and its powers, and they give non-zero contributions to the field in the laboratory frame of reference. For example, the term of order $1/c^4$ from (D2) give the following contribution to the part $\frac{1}{c} \frac{\partial {\bf A}}{\partial t}$ of the electric field
\begin{align}
-\frac{e}{6c^5} \frac{d^4}{dt^4}(R^2 {\bf v})= - \frac{2e}{3c^4} \dot{v}^2  \bm{\beta} -\frac{2e}{3c^3} (\bm{\beta}\cdot \ddot{{\bf v}}) \bm{\beta} - \frac{2}{3c^3} \beta^2 \ddot{{\bf v}} \nonumber \\
-\frac{e}{c^4} (\bm{\beta}\cdot \dot{{\bf v}}) \dot{{\bf v}} +\frac{e}{6c^4} ({\bf R} \cdot \bm{\beta}) \dot{\ddot{{\bf v}}}+\frac{e}{c^5} ({\bf R} \cdot \ddot{{\bf v}}) \dot{{\bf v}} \nonumber \\
+\frac{2e}{c^5} ({\bf R} \cdot \dot{{\bf v}}) \ddot{{\bf v}}- \frac{e}{6c^5} R^2 \ddot{\ddot{{\bf v}}}.
\end{align}
The first four terms from the r.h.s. of the above equations give contributions of order $R^0$ to the electric field of a point particle in the laboratory frame, contributions that are neglected in \cite{yag}. Similar considerations can be made regarding the truncation of the series (D1).

\end{document}